\documentclass[letterpaper]{article} 
\usepackage{aaai2027}  
\usepackage[hyphens]{url}  
\usepackage{graphicx} 
\usepackage{natbib}  
\usepackage{caption} 
\usepackage{booktabs}
\usepackage{colortbl}
\usepackage{arydshln}
\usepackage{amsmath}
\title{P-MUSE: Prompt-MIDI-Optional Model for Unified Instrumental Music Synthesis and Editing}

\author{
    Chong Jing\textsuperscript{\rm 1},
    Junan Zhang\textsuperscript{\rm 1},
    Jing Yang\textsuperscript{\rm 2},
    Yulun Wu\textsuperscript{\rm 2},
    Fan Fan\textsuperscript{\rm 2},
    Zhizheng Wu\textsuperscript{\rm 1}\corresponding
}

\affiliations{
    \textsuperscript{\rm 1}The Chinese University of Hong Kong, Shen Zhen\\
    \textsuperscript{\rm 2}Central Media Technology Institute, Huawei\\
    chongjing@link.cuhk.edu.cn, junanzhang@link.cuhk.edu.cn, yangjing201@huawei.com,\\
    wuyulun9@huawei.com, fanfan1@huawei.com, wuzhizheng@cuhk.edu.cn
}

\begin{document}

\maketitle

\begin{abstract}
MIDI-to-Music system renders the melody and rhythm of a target MIDI sequence into musical segment while cloning instrument timbre from a prompt recording.
Existing systems typically adopt one of two distinct paradigms: conditional generation with prompt audio alone, which remains applicable when aligned prompt MIDI is unavailable, and In-Context Learning with paired prompt audio and MIDI, which exploits cross-modal alignment for stronger control on MIDI following and timbre similarity.
We introduce P-MUSE, an instrumental MIDI-to-Music framework that unifies both paradigms via a multi-stage Curriculum-Learning supporting prompt-MIDI-optional inputs. P-MUSE further unifies music generation and local editing through a shared fill-in-the-middle formulation. Grounded in theoretical analysis and empirical study, we propose a phase-aware classifier-free guidance scheduling principle for Transcription-to-Audio systems, alongside a Tail-Drop strategy. Finally, to advance research in this field, we establish the first comprehensive benchmark, covering various prompt modes, generation/editing tasks, and four representative instruments: piano, guitar, bass, and drums. Demos are available at \url{https://p-muse.github.io/}.

\end{abstract}


\section{Introduction}
Transcription-to-Audio systems generate audio segments aligned with its transcriptions while preserving reference acoustic characteristics. As prominent instances of the Transcription-to-Audio system, MIDI-to-Music (MTM) renders target MIDI into musical segments in the timbre of prompt audio while Text-to-Speech (TTS) renders text into speech in the timbre of the target speaker.

Existing Transcription-to-Audio systems follow two complementary paradigms, depending on whether reference transcriptions are available: \emph{style prompting}~\cite{demerlé2024combiningaudiocontrolstyle,kim2025tokensynth,casanova2023yourttszeroshotmultispeakertts,wang2023neuralcodeclanguagemodels} conditions generation on target transcription (MIDI or Text) and prompt audio (Speech or Music) alone, offering robust and flexible inference when prompt transcription is unavailable; \emph{paired prompting}~\cite{tang2025midivalle,Cui_2026,jing2026anysynthzeroshotinstrumentcloningincontext,le2023voiceboxtextguidedmultilingualuniversal,eskimez2024e2ttsembarrassinglyeasy,chen2025f5ttsfairytalerfakesfluent,deng2025indexttsindustriallevelcontrollableefficient,du2025cosyvoice3inthewildspeech,du2024cosyvoicescalablemultilingualzeroshot} instead exploits aligned transcription--audio context to provide stronger transcription following and more precise timbre similarity.

Unifying these isolated paradigms into a single model is highly desirable: it grants the system universal applicability across heterogeneous inference scenarios and enables it to fully exploit all available reference information. Although recent TTS architectures have bridged analogous paradigms by co-training speaker encoders alongside generative backbones~\cite{hu2026qwen3ttstechnicalreport,zhang2025minimaxspeechintrinsiczeroshottexttospeech}, directly transferring this strategy to MTM encounters severe performance bottlenecks. Direct joint optimization of the encoder suffers because existing open-source instrumental music datasets lack timbre diversity. This challenge is further exacerbated in MTM, where musical timbres exhibit far more complex dynamic envelopes and richer harmonic overtones than human speech. This limitation raises a fundamental question: \textbf{How can MTM seamlessly bridge the two Transcription-to-Audio paradigms without being bottlenecked by the limited timbre diversity of current music datasets?}

\begin{figure}[t]
    \centering
    \includegraphics[width=\columnwidth]{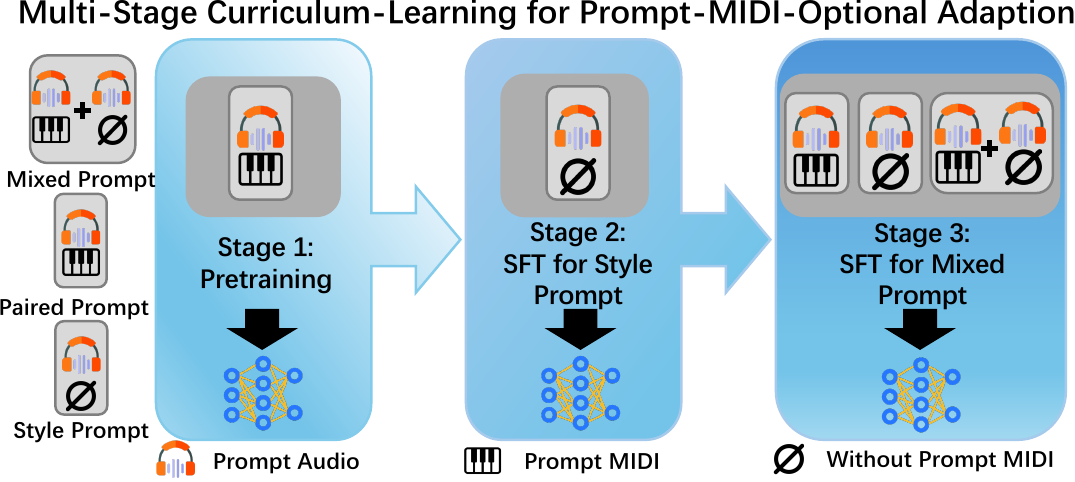}
    \caption{Multi-stage curriculum. Stage~1 learns paired-prompt In-Context Learning; Stage~2 adapts the model to style prompts through prompt-MIDI dropout; and Stage~3 trains for supporting mixed prompts while retaining compatibility with all three prompt modes.}
    \label{fig:multi_stage_overview}
\end{figure}

To address this challenge, we introduce \textbf{P-MUSE}, a \textbf{P}rompt-\textbf{M}IDI-\textbf{O}ptional Model for \textbf{U}nified \textbf{S}ynthesis and \textbf{E}diting in instrumental music. As illustrated in Figure~\ref{fig:multi_stage_overview}, P-MUSE adopts a progressive multi-stage Curriculum-Learning strategy~\cite{bengio2009curriculum,graves2017automated}. It inherits the strong MIDI following and timbre preservation priors established during paired In-Context Learning (ICL) pre-training, then progressively transfers these capabilities to the unpaired \emph{style-prompt} setting and ultimately the \emph{mixed-prompt} regime. This progressive capability transfer improves input robustness while maximizing the use of available reference information. Furthermore, P-MUSE adopts a fill-in-the-middle (FIM) formulation~\cite{bavarian2022efficienttraininglanguagemodels} to unify full-segment music generation and local editing under the same framework without changing the training objective.

Existing CFG schedules for Transcription-to-Audio systems are predominantly heuristic and lack clear theoretical interpretability. To address this limitation, we combine theoretical analysis with empirical observations to derive a phase-aware CFG scheduling principle for Transcription-to-Audio systems. Accordingly, we propose Tail-Drop, a simple yet effective strategy that disables CFG during the late stages of reverse integration. Tail-Drop not only preserves transcription following and timbre similarity capabilities while improving overall generation quality, but also cuts inference computational overhead. Extensive evaluations across both MTM and TTS demonstrate the strong cross-task generalizability and effectiveness of our approach.

To systematically evaluate P-MUSE, we construct a comprehensive benchmark covering three prompting settings: paired, style, and mixed. The benchmark also covers both full-segment generation and local editing tasks.

In summary, our main contributions are as follows:
\begin{itemize}
\item We propose P-MUSE, the first unified MIDI-to-Music framework that integrates full-segment generation and local editing across various instruments (guitar, piano, bass, and drums) via a shared fill-in-the-middle FIM objective.
\item We introduce a progressive multi-stage Curriculum-Learning strategy that transfers strong paired In-Context Learning priors to prompt-MIDI-optional scenarios.
\item We derive a phase-aware CFG scheduling principle for Transcription-to-Audio systems and introduce the Tail-Drop strategy. It enhances generation quality while maintaining strict transcription following and timbre similarity, demonstrating strong generalizability across both MTM and TTS tasks.
\end{itemize}
\section{Related Work}

\subsection{Zero-Shot Text-to-Speech}

Zero-shot TTS aims to customize a target speaker's voice with just a few seconds of a speech prompt. Some works treat the prompt as In-Context evidence, using paired speech and text to guide generation or editing~\cite{le2023voiceboxtextguidedmultilingualuniversal,eskimez2024e2ttsembarrassinglyeasy,chen2025f5ttsfairytalerfakesfluent,liu2024autoregressivediffusiontransformertexttospeech}. This formulation retains detailed acoustic and linguistic cues in the prompt, but generally requires that the prompt text is available.

Another line distills speaker identity into a dedicated representation through a speaker encoder, as in YourTTS and IndexTTS~\cite{casanova2023yourttszeroshotmultispeakertts,deng2025indexttsindustriallevelcontrollableefficient}. Such representations provide a compact and flexible conditioning interface, but discard part of the fine-grained in-context evidence. Recent systems including Qwen3-TTS, MiniMax-Speech, and X-Voice explore hybrid designs that combine in-context reference audio with speaker representations~\cite{hu2026qwen3ttstechnicalreport,zhang2025minimaxspeechintrinsiczeroshottexttospeech,xu2026xvoiceenablingspeak30}.

\subsection{MIDI-to-Music Synthesis}

Existing MTM systems likewise follow two prompt configurations. Some systems condition on style prompt without paired MIDI. CTD~\cite{demerlé2024combiningaudiocontrolstyle} disentangles timbre and melodic embeddings to guide conditional generation via diffusion models, and TokenSynth~\cite{kim2025tokensynth} leverages CLAP embeddings as timbre representations conditioned on REMI sequences~\cite{huang2020popmusictransformerbeatbased} for autoregressive music synthesis. Because they do not require aligned prompt MIDI, these systems remain applicable when only a reference recording is available.

In contrast, paired-prompt systems provide both prompt audio and its aligned MIDI. MIDI-VALLE uses paired audio-MIDI prompting for expressive piano synthesis~\cite{tang2025midivalle}, and Break-the-Beat! studies controllable MIDI-to-drum synthesis with paired symbolic and acoustic context~\cite{Cui_2026}. The aligned prompt exposes cross-modal correspondences between acoustic timbre and performance events, which can strengthen MIDI following and timbre similarity.
\section{Method}

\begin{figure*}[t]
    \centering
    \includegraphics[width=\textwidth]{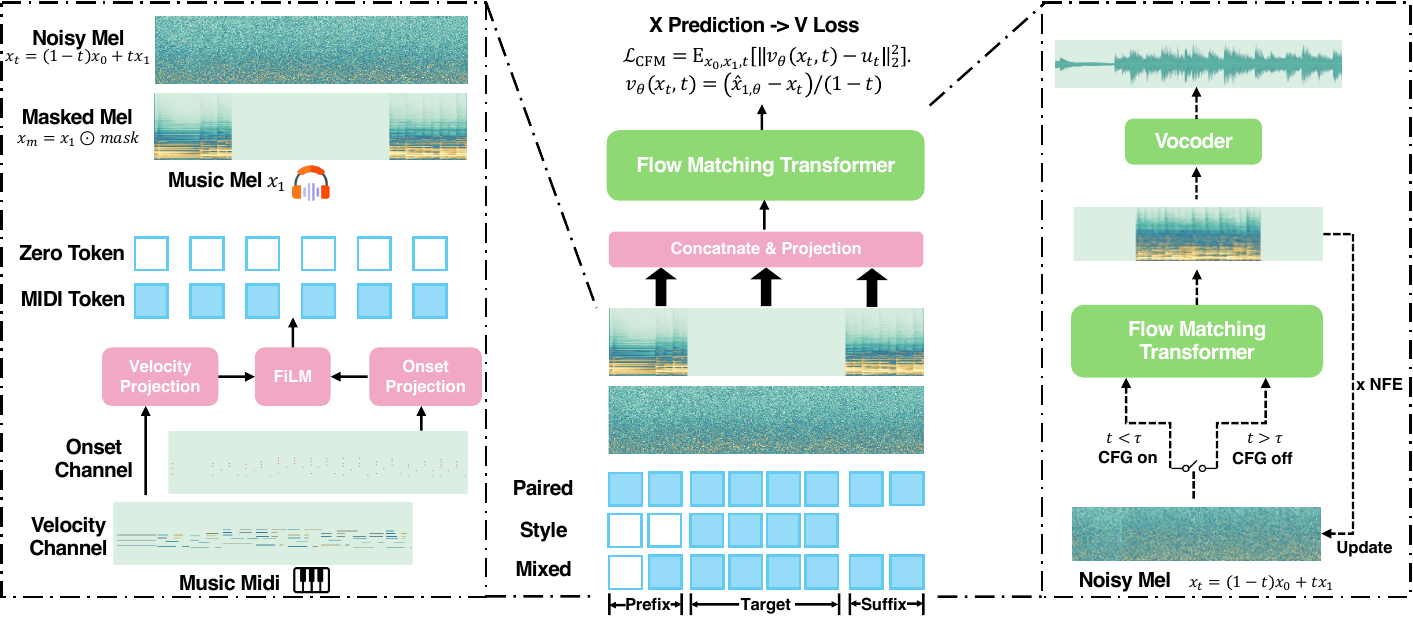}
    \caption{P-MUSE architecture. During training (left), frame-aligned MIDI, masked clean mel prompt, and noisy mel are fused and processed by the flow-matching Transformer with $x$-prediction and $v$-loss. During inference (right), Euler solver updates the inverse ODE process with Tail-Drop strategy before waveform reconstruction by the vocoder.}
    \label{fig:model_overview}
\end{figure*}

\begin{figure}[t]
    \centering
    \includegraphics[width=\columnwidth]{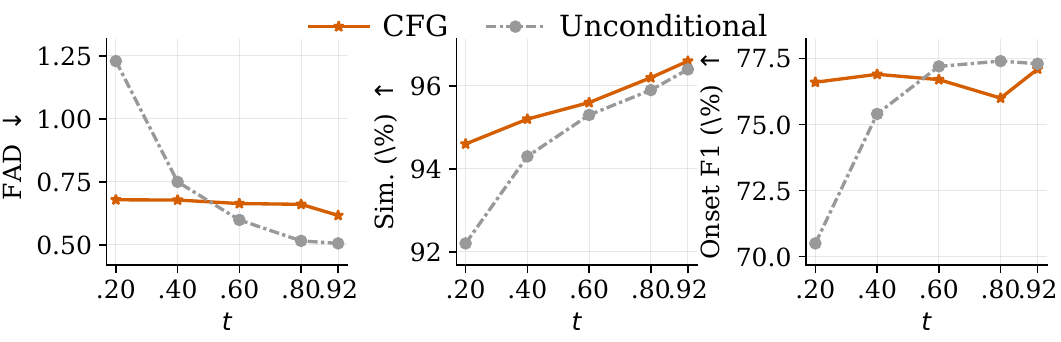}
    \caption{Probe on generation conducted on a fixed evaluation set $\{x^i\}_{i=1}^{4000}$. Each curve starts from the real-path latent $x_t^{\mathrm{real, i}}=(1-t)x_0+t x_1^i$ at the indicated time and integrates back to the data endpoint with $v_{\mathrm{CFG}}(x_t,t)$ or $v_\theta(x_t,t,\emptyset)$.}
    \label{fig:refinement_probe}
\end{figure}

\subsection{Preliminaries}
\label{sec:Preliminaries}
\subsubsection{Conditional Flow Matching}

We formulate MTM generation as transport from a simple prior distribution $p_{\mathrm{prior}}$ to the data distribution $p_{\mathrm{data}}$. Let $p_t$, $t\in[0,1]$, denote the intermediate marginal distributions, with $p_0=p_{\mathrm{prior}}$ and $p_1=p_{\mathrm{data}}$. Their probability-flow ODE (PF-ODE) is governed by the continuity equation:
\begin{equation}
    \partial_t p_t(x)+\nabla_x\cdot\bigl(u_t(x)p_t(x)\bigr)=0,
\end{equation}
where $u_t$ is the velocity field of the marginal transport. Because $u_t$ is intractable, conditional flow matching (CFM)~\cite{lipman2023flowmatching} instead conditions the path on a data sample $x_1\sim p_{\mathrm{data}}$ and uses a tractable conditional velocity. Its marginal field is $u_t(x)=\mathrm{E}_{x_1\mid x_t}[u_t(x_t\mid x_1)\mid x_t]$, and the resulting objective has the same gradients as direct flow matching~\cite{mathieu2024flow}.

Specifically, we sample $x_0\sim p_{\mathrm{prior}}=\mathcal{N}(0,I)$ independently of $x_1$. A conditional path can be constructed as
\begin{equation}
    x_t=\alpha_t x_1+\sigma_t x_0,
\end{equation}
with conditional velocity $u_t(x_t\mid x_1)=\dot{\alpha}_t x_1+\dot{\sigma}_t x_0$. P-MUSE uses the rectified linear schedule $(\alpha_t,\sigma_t)=(t,1-t)$ for $t\sim\mathcal{U}[0,1]$~\cite{liu2022flowstraightfastlearning}, yielding $x_t=(1-t)x_0+t x_1$ and $u_t(x_t\mid x_1)=x_1-x_0$. CFM trains the neural field $v_\theta$ with
\begin{equation}
    \mathcal{L}_{\mathrm{CFM}}=
    \mathrm{E}_{t,x_0,x_1}
    \left[\left\|v_\theta(x_t,t)-u_t(x_t\mid x_1)\right\|_2^2\right].
\end{equation}

\subsubsection{Closed-Form Conditional Velocity Field}
\label{sec:closed_form_cfm}

Following analysis of former literatures~\cite{liu2025navigation,bertrand2025closedformflowmatchinggeneralization,Biroli_2024,gao2024flowmatchingmodelsmemorize,kamb2025analytictheorycreativityconvolutional}, with a finite closed dataset $\mathcal{D}=\{{x_1^i\}}_{i=1}^{N}$, a Gaussian prior distribution and a linear interpolation probability path, $u^\star(x_t,t):=\mathrm{E}_{x_1\mid x_t}[u_t(x_t\mid x_1)\mid x_t]$ admits a closed expression for $t\in [0,1]$:
\begin{equation}
u_{\mathrm{uncond}}^\star(x_t,t)=
\frac{\sum_{i=1}^{N}\gamma_i^{\emptyset}(x_t,t)x_1^i-x_t}{1-t},
\label{eq:unconditional_oracle}
\end{equation}
where
\begin{equation}
\gamma_i^{\emptyset}(x_t,t)=
\frac{\exp\!\left(-\frac{\|x_t-tx_1^i\|_2^2}{2(1-t)^2}\right)}
{\sum_{j=1}^{N}\exp\!\left(-\frac{\|x_t-tx_1^j\|_2^2}{2(1-t)^2}\right)}.
\end{equation}
Thus,  $u^\star(x_t,t)$ averages the velocities associated with all samples, weighted by their posterior weights given $x_t$.

\subsubsection{Classifier-Free Guidance}

Classifier-free guidance (CFG) combines the full conditional field with a contrast against the dropped-condition field:
\begin{equation}
\begin{array}{rcl}
v_{\mathrm{CFG}}(x_t,t,c) &=& v_\theta(x_t,t,c) \\
&+& s_{\mathrm{cfg}}\left[v_\theta(x_t,t,h)-v_\theta(x_t,t,\emptyset)\right],
\end{array}
\label{eq:standard_cfg}
\end{equation}
where $s_{\mathrm{cfg}}$ is the CFG scale.

\subsection{Architecture}

\subsubsection{Overview}

Figure~\ref{fig:model_overview} summarizes the shared training and inference pipeline under three prompt settings. P-MUSE unifies MTM generation and editing through a fill-in-the-middle (FIM) formulation~\cite{bavarian2022efficienttraininglanguagemodels}, using a Flow Matching Transformer as its generative backbone.

\subsubsection{FIM Construction}

For paired music-MIDI example, we sample prefix, middle, and suffix regions after batching. The middle split is the target we want to generate or edit while the prefix and suffix remain clean context for reference. Therefore, generation and editing are represented as completing a target region given the constructed context so both tasks are able to share the same model and training objective.

\subsubsection{Generative Backbone}

We extract mel-spectrograms as low-dimensional representation in latent space~\cite{kingma2022autoencodingvariationalbayes}. For MIDI conditioning, P-MUSE rasterizes single-track MIDI into a two-channel piano roll~\cite{hawthorne2018onsetsframesdualobjectivepiano,kim2018neuralmusicsynthesisflexible}: velocity channel captures note content and performance-dynamics information, while the onset channel carries onset and sustain information. Two independent Multi-layer Perceptron (MLP) first map the velocity and onset channels to continuous features. We then apply Feature-wise Linear Modulation (FiLM)~\cite{perez2017filmvisualreasoninggeneral} to modulate the velocity features with onset features. The resulting MIDI tokens retain structural content while making its representation sensitive to dynamic cues. Unlike text in TTS, each MIDI note is explicitly and naturally timestamped. We therefore set the piano-roll frame rate to match the mel-spectrogram hop size, yielding frame-level cross-modal alignment. 

At each frame, the model concatenates three feature groups along the feature dimension: the MIDI tokens, masked clean mel prompt, and noisy mel state $x_t$. A linear projection then maps the fused feature to the hidden dimension of a Flow Matching Transformer based on Diffusion Transformer (DiT) \cite{peebles2023dit}. 

\subsection{Multi-Stage Curriculum-Learning}

\begin{table}[t]
    \centering
    {\small
    \setlength{\tabcolsep}{1.5pt}
    \begin{tabular}{lrrr}
    \toprule
    Stage~1 inference & FAD $\downarrow$ & Sim. (\%) $\uparrow$ & F1 (\%) $\uparrow$ \\
    \midrule
    With prompt MIDI & 0.773 & 93.5 & 73.3 \\
    Without prompt MIDI & 0.855 & 90.2 & 68.6 \\
    \bottomrule
    \end{tabular}
    }
    \caption{OOD stress test on the model after Stage~1. Prompt MIDI is dropped only at inference.}
    \label{tab:prompt_probe}
\end{table}

Figure~\ref{fig:multi_stage_overview} summarizes the progression from paired to style and mixed prompt setting. 
Specifically, Stage~2 and~3 perform supervised fine-tuning (SFT) on a high-quality subset.
Each stage initializes from the preceding checkpoint, progressively extending the input interface without changing the shared model architecture.

\subsubsection{Stage 1: paired-prompt pretraining.}
We begin to train P-MUSE based on In-Context Learning with paired prompt, which supplies the strong performance in MIDI following and timbre preservation. During CFG, when the MIDI-dropped branch is triggered, all MIDI tokens will be substituted by zero-valued tensors.

The CFG formulation in Stage~1 makes style prompting a natural adaptation target: A style prompt removes only prompt MIDI while retaining target MIDI; it is thus a structured special case of the MIDI-dropped branch. 
Specifically, the dropout in Stage~1 equips the model with two fundamental denoising abilities under two conditions: full or with no MIDI. Learning to denoise under style prompt—where the model relies exclusively on the target MIDI—is therefore not forcing the model to master a completely new ability from scratch. \textbf{Instead, it represents a natural interpolation between these two pre-established abilities.}
\subsubsection{Out-of-Distribution Stress Test}
We explore this transfer feasibility with an inference-time out-of-distribution (OOD) stress test with the checkpoint of Stage~1 by removing prompt MIDI from paired prompt inputs. Table~\ref{tab:prompt_probe} shows a measurable degradation but no performance collapse, indicating the feasibility of further Curriculum-Learning stages dedicated for prompt-MIDI-optional adaptation.

\subsubsection{Stage 2: style-prompt adaptation.}
Stage~2 uses prefix-middle FIM setting, in which a visible prefix precedes the target middle region and no suffix is provided. When the MIDI-dropped branch is triggered, we further either drop prompt MIDI or drop full MIDI with a rate of 0.5.

\subsubsection{Stage 3: mixed-prompt training.}
The final stage extends the input to not only previous two prompt settings, but also mixed prompt so that the model can exploit prompt information as fully as possible. We retain the standard FIM context and prepend the style prompt. The resulting input combines an audio-only reference with paired FIM context and target MIDI, teaching a single model to use both sources without changing its interface. After the two-stage SFT, the resulting checkpoint is compatible with all three input settings.

\subsection{Phase-Aware CFG Schedule.}
\label{sec:tail_drop_cfg}

Prior work shows that there is an optimal interval (e.g. $t\in[0.1,0.6]$) to apply CFG in image generation task conditioned on class labels~\cite{kynkaanniemi2024limitedguidance,liu2025navigation}. Whether this principle transfers to Transcription-to-Audio systems like MTM and TTS remains unclear, because guidance must additionally preserve transcription following and timbre similarity.

The closed-form analysis above concerns only $u_{\mathrm{uncond}}^\star$. We extend it to conditional velocity fields under CFG. This extension exposes an important difference between class-level and instance-level conditions. Denoting a subset of the dataset with a class-level condition $c$ as $\mathcal{I}_c=\{i:c_i=c\}$, the closed-form of $u^\star(x_t,t,c):=\mathrm{E}_{x_1\mid x_t,c}[u_t(x_t\mid x_1)\mid x_t,c]$ equals
\begin{equation}
u_{\mathrm{cond}}^\star(x_t,t,c)=
\frac{\sum_{i\in\mathcal{I}_c}\gamma_i^c(x_t,t)x_1^i-x_t}{1-t},
\label{eq:label_cfg_u}
\end{equation}
where
\begin{equation}
\gamma_i^c(x_t,t)=
\frac{\exp\!\left(-\frac{\|x_t-tx_1^i\|_2^2}{2(1-t)^2}\right)}
{\sum_{j\in\mathcal{I}_c}\exp\!\left(-\frac{\|x_t-tx_1^j\|_2^2}{2(1-t)^2}\right)}.
\end{equation}
Thus, class-level conditioning retains posterior weights averaging within the selected class in the early trajectory but will collapse to one-hot in the late trajectory. 

At the other extreme, an instance-level condition uniquely identifies $x_{k(c)}$. Accordingly, its posterior weights remain one-hot across the entire trajectory, and the conditional velocity field degenerates to
\begin{equation}
u_{\mathrm{cond}}^\star(x_t,t,c)=
\frac{x_{k(c)}-x_t}{1-t}=x_{k(c)}-x_0.
\label{eq:instance_cfg_u}
\end{equation}
The key insight from Eq.~\ref{eq:instance_cfg_u} is that instance-level conditions yield a velocity field consistently dominated by a single target sample. In Transcription-to-Audio systems, conditions (detailed transcriptions + timbre features) are far more specific than sparse class labels, making their conditional fields behave closer to instance-level than class-level conditioning. This implies that once the trajectory has accumulated sufficient structural information, the unconditional contrast offered by CFG may introduce over-saturation artifacts rather than corrective guidance. We verify this hypothesis through an inference-time probe. We first construct a testset with 4000 samples $\{x^i\}_{i=1}^{4000}$. Given the ground-truth target latent $x_1^i$ and a fixed noise sample $x_0$, we construct $x_t^{\mathrm{real, i}}=(1-t)x_0+t x_1^i$, then integrate the model-predicted field from chosen start t with $v_{\mathrm{CFG}}(x_t,t)$ or $v_\theta(x_t,t,\emptyset)$. Figure~\ref{fig:refinement_probe} shows that the dropping guidance outperforms in FAD and onset F1 after t=0.6, with only a slight timbre similarity difference.

These insights naturally reveal two distinct, phase-dependent regimes during the sampling trajectory. In the \textbf{Structure-Formation Phase} ($0 \le t < \tau$), the reverse process focuses on recovering core structural information dictated by the conditions. While applying early CFG over an uncollapsed posterior in class-level generation magnifies noise variance and induces trajectory drift, instance-level conditioning in Transcription-to-Audio systems collapses the posterior to $x_{k(c)}$ from the very beginning, making early guidance both safe and vital for establishing these structural foundations. In the subsequent \textbf{Detail-Refinement Phase} ($\tau \le t \le 1$), the sampling trajectory transitions to optimizing fine-grained details; dropping CFG at this stage prevents over-steering, over-saturation, and memorization artifacts, thereby enhancing overall generation quality.

Grounded in this phase-aware principle, we introduce Tail-Drop, a simple yet effective classifier-free guidance schedule:
\begin{equation}
    v_{\mathrm{Tail\text{-}Drop}}(x_t,t,c) = \begin{cases} 
    v_{\mathrm{CFG}}(x_t,t), & 0 \le t < \tau \\
    v_\theta(x_t,t,\emptyset), & \tau \le t \le 1 
    \end{cases}
\end{equation}
where $\tau$ denotes the phase-transition threshold separating conditional structure formation from detail refinement.
\section{Experiments}
\begin{table}[t]
    \centering
    {\small
    \setlength{\tabcolsep}{1.5pt}
    \begin{tabular}{@{}lllrrr@{}}
    \toprule
    Model & Prompt & Track & FAD $\downarrow$ & Sim. (\%) $\uparrow$ & Onset F1 (\%) $\uparrow$ \\
    \midrule
    P-MUSE & Paired & Piano & \textbf{1.238} & \textbf{92.8}$\pm0.8$ & \textbf{69.9}$\pm3.0$ \\
    MIDI-VALLE & Paired & Piano & 7.655 & 84.3$\pm1.4$ & 54.9$\pm2.8$ \\
    \addlinespace[1pt]
    \hdashline
    \addlinespace[1pt]
    P-MUSE & Style & Bass & \textbf{0.560} & \textbf{93.3}$\pm0.8$ & \textbf{73.6}$\pm4.2$ \\
    CTD & Style & Bass & 4.031 & 80.9$\pm1.7$ & 50.5$\pm4.8$ \\
    TokenSynth & Style & Bass & 1.789 & 85.6$\pm1.5$ & 38.1$\pm4.3$ \\
    P-MUSE & Style & Guitar & \textbf{1.058} & \textbf{92.3}$\pm0.9$ & \textbf{67.1}$\pm4.6$ \\
    CTD & Style & Guitar & 5.428 & 82.7$\pm1.3$ & 49.9$\pm4.2$ \\
    TokenSynth & Style & Guitar & 2.477 & 86.4$\pm1.2$ & 36.9$\pm3.5$ \\
    P-MUSE & Style & Piano & \textbf{1.700} & \textbf{90.9}$\pm1.2$ & \textbf{66.7}$\pm2.9$ \\
    CTD & Style & Piano & 6.603 & 82.9$\pm1.5$ & 47.2$\pm3.4$ \\
    TokenSynth & Style & Piano & 3.229 & 86.3$\pm1.3$ & 39.4$\pm3.1$ \\
    \bottomrule
    \end{tabular}
    }
    \caption{Track-level generation comparison with available baselines. Results are comparable only within each prompt-mode and instrument-track group; bold indicates the best result in the group.}
    \label{tab:baseline_results}
\end{table}

\begin{table}[t]
    \centering
    {\small
    \setlength{\tabcolsep}{2.0pt}
    \begin{tabular}{@{}lrrrrrr@{}}
    \toprule
    & \multicolumn{3}{c}{\textbf{Generation}} & \multicolumn{3}{c}{\textbf{Editing}} \\
    \cmidrule(lr){2-4}\cmidrule(l){5-7}
    Prompt & FAD$\downarrow$ & Sim.(\%)$\uparrow$ & F1(\%)$\uparrow$ & FAD$\downarrow$ & Sim.(\%)$\uparrow$ & F1(\%)$\uparrow$\\
    \midrule
    Paired & 0.602 & 94.1$\pm.4$ & 74.8$\pm1.9$ & 0.835 & 93.1$\pm.2$ & 71.8$\pm.8$ \\
    Style  & 0.558 & 92.9$\pm.5$ & 73.0$\pm2.0$ & 0.808 & 92.3$\pm.2$ & 74.4$\pm.7$ \\
    Mixed  & 0.668 & 94.2$\pm.4$ & 74.2$\pm2.0$ & 0.831 & 93.0$\pm.2$ & 71.8$\pm.8$ \\
    \bottomrule
    \end{tabular}
    }
    \caption{Overall P-MUSE results across the three benchmark settings, averaged over piano, guitar, bass, and drums. F1 denotes onset F1.}
    \label{tab:overall_results}
\end{table}

\begin{table}[t]
    \centering
    {\small
    \setlength{\tabcolsep}{3.5pt}
    \begin{tabular}{lrrr}
    \toprule
    Prompt & FAD $\downarrow$ & Sim. (\%) $\uparrow$ & Onset F1 (\%) $\uparrow$ \\
    \midrule
    Mixed & \textbf{0.592} & \textbf{93.6}$\pm.2$ & \textbf{75.5}$\pm1.1$ \\
    Paired-only & 0.679 & 92.0$\pm.4$ & 73.2$\pm1.3$ \\
    Style-only & 0.641& 92.9$\pm.3$ & 73.6$\pm1.2$ \\
    \bottomrule
    \end{tabular}
    }
\caption{Performance comparison of three prompt settings under information bottleneck.}
\label{tab:context_bottleneck}
\end{table}

\begin{table*}[t]
    \centering
    {\small
    \setlength{\tabcolsep}{3.0pt}
    \begin{tabular}{lrrrrrr}
    \toprule
    & \multicolumn{2}{c}{\textbf{Paired Prompt}} & \multicolumn{3}{c}{\textbf{Style Prompt}} & \textbf{Mixed Prompt} \\
    \cmidrule(lr){2-3}\cmidrule(lr){4-6}\cmidrule(l){7-7}
    Metric & P-MUSE & MIDI-VALLE & P-MUSE & CTD & TokenSynth & P-MUSE \\
    \midrule
    Audio Quality $\uparrow$ & \textbf{4.175} & 3.155 & \textbf{3.733} & 2.992 & 3.417 & 3.826 \\
    MIDI Following $\uparrow$ & \textbf{4.375} & 3.540 & \textbf{4.125} & 2.767 & 3.542 & 4.012 \\
    Timbre Similarity $\uparrow$ & \textbf{4.225} & 3.216 & \textbf{3.783} & 2.217 & 3.008 & 4.127 \\
    \bottomrule
    \end{tabular}
    }
    \caption{MOS generation comparison with available baselines across prompt settings. Ten listeners rate each aspect on a 1--5 scale; higher is better. Bold denotes the best result within each applicable prompt setting.}
    \label{tab:mos_generation}
\end{table*}

\begin{table}[t]
    \centering
    {\small
    \setlength{\tabcolsep}{5.0pt}
    \begin{tabular}{lrrr}
    \toprule
    Metric & Paired & Style & Mixed \\
    \midrule
    Audio Quality $\uparrow$ & 4.525 & 4.228 & 3.926 \\
    MIDI Following $\uparrow$ & 4.475 & 4.000 & 4.221 \\
    Timbre Similarity $\uparrow$ & 4.460 & 4.097 & 3.960 \\
    Context Coherence $\uparrow$ & 4.420 & 4.062 & 4.032 \\
    Transition Smoothness $\uparrow$ & 4.450 & 3.986 & 4.032 \\
    \bottomrule
    \end{tabular}
    }
    \caption{MOS results for P-MUSE local editing across prompt settings.}
    \label{tab:mos_editing}
\end{table}

\begin{table}[t]
    \centering
    {\small
    \setlength{\tabcolsep}{3.5pt}
    \begin{tabular}{@{}l@{\hspace{3.5pt}}lrrr@{}}
    \toprule
    Curriculum & Prompt & FAD $\downarrow$ & Sim. (\%) $\uparrow$ & Onset F1 (\%) $\uparrow$ \\
    \midrule
    S1$\rightarrow$S2$\rightarrow$S3 & Paired & \textbf{0.602} & \textbf{94.1}$\pm.4$ & 74.8$\pm1.9$ \\
    S1 & Paired & 0.773 & 93.5$\pm.4$ & 73.3$\pm2.0$ \\
    S1$\rightarrow$S2 & Paired & 0.686 & 93.7$\pm.4$ & 75.5$\pm1.9$ \\
    S1$\rightarrow$S3 & Paired & 0.715 & 93.9$\pm.4$ & \textbf{75.7}$\pm1.9$ \\
    Direct & Paired & 0.660 & \textbf{94.1}$\pm.4$ & 73.8$\pm1.9$ \\
    \addlinespace[1pt]
    \hdashline
    \addlinespace[1pt]
    S1$\rightarrow$S2$\rightarrow$S3 & Style & \textbf{0.558} & \textbf{92.9}$\pm.5$ & \textbf{73.0}$\pm2.0$ \\
    S1$\rightarrow$S2 & Style & 0.614 & 91.9$\pm.6$ & 72.3$\pm2.0$ \\
    S1$\rightarrow$S3 & Style & 0.717 & 92.5$\pm.5$ & 72.5$\pm2.0$ \\
    Direct & Style & 0.668 & 92.6$\pm.5$ & 71.0$\pm2.1$ \\
    \addlinespace[1pt]
    \hdashline
    \addlinespace[1pt]
    S1$\rightarrow$S2$\rightarrow$S3 & Mixed & 0.668 & \textbf{94.2}$\pm.4$ & 74.2$\pm2.0$ \\
    S1$\rightarrow$S3 & Mixed & 0.735 & 94.0$\pm.4$ & \textbf{75.2}$\pm1.9$ \\
    Direct & Mixed & \textbf{0.650} & 94.1$\pm.4$ & 73.6$\pm2.1$ \\
    \bottomrule
    \end{tabular}
    }
    \caption{Ablation for the multi-stage Curriculum-Learning.}
    \label{tab:stage_ablation}
\end{table}

\subsection{Dataset}

We construct the training dataset from open-source and high-quality aligned MIDI-audio datasets: Slakh\citep{manilow2019slakh}, SynthTab\citep{zang2024synthtab}, the Expanded Groove MIDI Dataset (e-gmd)\citep{callender2020egmd}, Guitarset~\cite{xi2018guitarset} and Maestro~\cite{hawthorne2018enabling}.
We also incorporate The Neural Audio Synthesis (NSynth) dataset~\cite{engel2017nsynth}, a large-scale collection of over 300,000 annotated isolated musical notes, to render single-track MIDI sequences separated from Lakh dataset~\cite{raffel2016learning} by General MIDI rules.
The original NSynth dataset is first upsampled to 48 kHz from 16 kHz using the LavaSR~\footnote{\url{https://huggingface.co/YatharthS/LavaSR}}.
Finally, the training dataset contains 3.44M audio-MIDI clips including piano, guitar, bass and drums, covering 216 distinct timbres with a total duration of 10,441 hours. On average, each timbre contributes around 50 hours of training data.

\subsection{Benchmark}

We construct three benchmark variants corresponding to paired, style, and mixed prompt settings. All test instruments are held out from the training set. For each instrument family (bass, drums, guitar, piano), we sample 100 target MIDI-audio pairs. Each pair yields one generation example and five editing examples (note addition, note deletion, pitch shift, velocity scaling, timing perturbation), resulting in 600 examples per instrument family(100 generation + 500 editing) and 2,400 examples per benchmark variant.

The benchmark is available at P-MUSE-eval~\footnote{\url{https://github.com/FEAfeatherTHER/P-MUSE-eval}}.

\subsection{Metrics}

We report Fr\'echet Audio Distance (FAD) with VGGish Features~\cite{hershey2017cnnarchitectureslargescaleaudio} for audio quality \cite{kilgour2019fad}, similarity (Sim.) for timbre consistency \cite{shi2022instrumentembedding}, and onset F1 for MIDI following. Lower FAD is better; higher similarity and onset F1 are better. For similarity and onset F1, we report the sample mean and 95\% bootstrap confidence-interval half-width (10,000 resamples). For onset F1, we transcribe both the generated audio and its corresponding ground-truth audio with YourMT3~\cite{chang2024yourmt3multiinstrumentmusictranscription}, and compare the resulting MIDI transcriptions rather than directly comparing a transcription with the source MIDI. This protocol reduces the effect of systematic transcription error. Additionally, for editing, FAD is evaluated over the regenerated target region together with a one-second context window on each side, which additionally tests boundary continuity; similarity is computed against the original prompt.

We further conduct a Mean Opinion Score (MOS) listening test with 10 listeners. For the generation task, samples are evaluated in a randomized and blind setup. Listeners rate each sample on a five-point scale for audio quality, MIDI following, and timbre similarity. For local editing, due to the absence of baseline comparisons, we perform an absolute MOS evaluation; listeners additionally rate context coherence, which measures consistency between the regenerated region and its retained context, and transition smoothness, which measures the naturalness of the edit boundaries.

\subsection{Main Results}

We first compare P-MUSE with available baselines in matched generation settings (Table~\ref{tab:baseline_results}). MIDI-VALLE is designed for piano performance rendering with paired prompt. CTD and TokenSynth supports style prompting but excludes drums. No open-source baseline supports drum synthesis, so drums are reported only in P-MUSE's overall benchmark results. None of the baselines supports P-MUSE's mixed context or local editing. In contrast, P-MUSE supports paired, style, and mixed prompts for piano, guitar, bass, and drums. The comparisons are made within the same prompt mode and instrument track. Baseline evaluations follow each model's supported prompt mode and publicly available implementation; detailed inference settings are provided in the supplementary material.

P-MUSE outperforms every available baseline in every comparable instrument-track group. On paired-prompt piano, it delivers a substantial advantage over MIDI-VALLE. Across all three style-prompt tracks, P-MUSE consistently surpasses the strongest baselines by a substantial margin, achieving lower FAD alongside higher timbre similarity and onset F1 scores. These per-track results establish a consistent advantage, while P-MUSE additionally supports drums, mixed prompting, and local editing, which are unavailable in the compared baselines.

Table~\ref{tab:overall_results} further reports the results averaged over piano, guitar, bass and drums on all three benchmarks. These results demonstrate the performance robustness of P-MUSE under flexible prompt-MIDI-optional settings, indicating its strong practical utility in real-world applications.

Table~\ref{tab:overall_results} only demonstrates the prompt-setting compatibility, whether the mixed-prompt mode enables the model to fully leverage information from both paired and style prompt to enhance overall performance remains unclear.

To systematically investigate this hypothesis, we construct a special testset where the paired prompt is truncated to at most 3 seconds, creating a severe information bottleneck, while an audio-only style prompt is provided alongside it. In this setting, the Mixed setting feeds both the 3-second paired prompt and the full style prompt to the model simultaneously, whereas Paired-only and Style-only baselines are strictly restricted to their respective single prompt components.
 
As reported in Table~\ref{tab:context_bottleneck}, the Mixed prompt mode achieves superior performance across all metrics compared to both single-prompt baselines. Specifically, it reduces FAD by $12.8\%$ and $7.6\%$ relative to the Paired-only and Style-only baselines, respectively. Furthermore, it yields consistent relative gains in both Onset F1 (up to $3.1\%$ over Paired-only and $2.6\%$ over Style-only) and timbre similarity. This confirms that P-MUSE effectively fuses complementary information from both prompt streams to overcome the severe information bottleneck.

Our human MOS evaluation results in Tables~\ref{tab:mos_generation} and~\ref{tab:mos_editing} across both generation and local editing tasks. For generation, the high MOS together with objective metrics further proves the advantages of P-MUSE on MTM. For local editing, P-MUSE also demonstrates stable performance across the core metrics, while additionally achieving high scores in Context Coherence and Transition Smoothness. This demonstrates that the FIM formulation enables the model to effectively leverage surrounding unmasked context, ensuring seamless temporal boundaries and holistic consistency between regenerated regions and retained context.

\begin{table}[t]
    \centering
    {\small
    \setlength{\tabcolsep}{3.5pt}
    \begin{tabular}{lrrr}
    \toprule
    CFG $\mathcal{I}$ & FAD $\downarrow$ & Sim. (\%) $\uparrow$ & Onset F1 (\%) $\uparrow$ \\
    \midrule
    full & 0.602 & \textbf{94.1}$\pm.4$ & 74.8$\pm1.9$ \\
    $[0.00,0.60]$ & \textbf{0.516} & 93.7$\pm.4$ & \textbf{76.0}$\pm1.8$ \\
    $[0.00,0.80]$ & 0.541 & 93.9$\pm.4$ & 75.4$\pm1.8$ \\
    $[0.00,0.92]$ & 0.523 & 94.0$\pm.4$ & 75.2$\pm1.8$ \\
    $[0.04,1.00]$ & 0.717 & 92.4$\pm.5$ & 71.9$\pm1.9$ \\
    $[0.08,1.00]$ & 1.499 & 90.6$\pm.6$ & 68.9$\pm2.1$ \\
    \bottomrule
    \end{tabular}
    }
    \caption{Ablations of Tail-Drop on P-MUSE.}
    \label{tab:tail_drop_grid}
\end{table}
    
\begin{table}[t]
    \centering
    {\small
    \setlength{\tabcolsep}{3.5pt}
    \begin{tabular}{llrrr}
    \toprule
    CFG $\mathcal{I}$ & Lang. & FSD $\downarrow$ & Sim. (\%) $\uparrow$ & WER (\%) $\downarrow$ \\
    \midrule
    full & En & 0.207 & 62.4$\pm.7$ & 1.99$\pm.33$ \\
    $[0.00,0.60]$ & En & 0.197 & 61.9$\pm.7$ & 2.00$\pm.33$ \\
    $[0.00,0.80]$ & En & \textbf{0.185} & \textbf{62.6}$\pm.7$ & \textbf{1.95}$\pm.33$ \\
    $[0.06,1.00]$ & En & 0.304 & 57.2$\pm.8$ & 2.32$\pm.33$ \\
    \addlinespace[1pt]
    \hdashline
    \addlinespace[1pt]
    full & Zh & 0.556 & 65.6$\pm.5$ & 7.19$\pm.65$ \\
    $[0.00,0.60]$ & Zh & 0.541 & 65.0$\pm.5$ & \textbf{7.12}$\pm.65$ \\
    $[0.00,0.80]$ & Zh & \textbf{0.538} & \textbf{65.7}$\pm.5$ & 7.20$\pm.66$ \\
    $[0.06,1.00]$ & Zh & 0.622 & 62.3$\pm.6$ & 7.26$\pm.66$ \\
    \bottomrule
    \end{tabular}
    }
    \caption{Ablations of Tail-Drop on F5-TTS.}
    \label{tab:tail_drop_tts}
\end{table}

\subsection{Ablation on Multi-Stage Curriculum-Learning}

We further examine how each stage of the multi-stage Curriculum-Learning training strategy contributes to this prompt-mode compatibility.

To isolate and evaluate the efficacy of each training stage, we construct a series of stage-wise baseline variants.
\begin{itemize}
    \item S1 variant is the model after Stage~1 pretraining.
    \item The S1$\rightarrow$S2 variant stops before S3, testing whether Stage~3 degrades the paired and style-prompt capabilities established earlier.
    \item S1$\rightarrow$S3 variant bypasses Stage~2, testing whether explicit style-prompt adaptation is necessary before joint training.
    \item Direct S3 variant trains on the prompt settings of Stage~3 from scratch for 418,000 steps to isolate the value of the multi-stage Curriculum-Learning.
\end{itemize}    

Table~\ref{tab:stage_ablation} reports the generation results; the corresponding editing results are provided in the supplementary material.
Relative to the S1 variant, the full Curriculum-Learning enhances all evaluation metrics under paired prompts while successfully enabling prompt-MIDI-optional capabilities.
Compared with S1$\rightarrow$S2, the full Curriculum-Learning lowers FAD and improves similarity on paired and style prompts, showing that Stage~3 maintains the previously supported settings despite a slight trade-off in onset F1.
Conversely, bypassing Stage~2 (S1$\rightarrow$S3) degrades FAD and similarity across the three prompt modes. Stage~2 is therefore important for transferring the paired In-Context Learning prior to audio-only style prompts before Stage~3. 
Direct S3 without curriculum yields lower FAD under paired and mixed prompt settings, but suffers noticeable degradation in onset F1 accuracy. Overall, these results validate that the progressive multi-stage curriculum achieves smooth and robust capability transfer without sacrificing generation fidelity or context adaptability.

\subsection{Ablation on Tail-Drop Strategy}
To evaluate the generalizability of the Tail-Drop strategy across Transcription-to-Audio systems, we assess its performance on both MTM using P-MUSE and TTS using F5-TTS~\cite{chen2025f5ttsfairytalerfakesfluent}.
For P-MUSE, we report the paired-prompt generation ablation in Table~\ref{tab:tail_drop_grid}; all editing results and the style- and mixed-prompt results are provided in the supplementary material.
For F5-TTS, we test on the English and Chinese Seed-TTS test sets with $32$ Euler steps and a CFG scale of $2.0$, as shown in Table~\ref{tab:tail_drop_tts}. Following the official \texttt{seed-tts-eval} pipeline\footnote{\url{https://github.com/BytedanceSpeech/seed-tts-eval}}, we compute Sim and WER, and calculate FSD using WavLM Base Plus~\cite{kim2026understandingfrechetspeechdistance,Chen_2022}.
Across all experiments, each interval $\mathcal{I}$ denotes the normalized time range where standard CFG is active; the model switches to the unconditional field outside $\mathcal{I}$.

Across both tasks, the empirical results consistently validate the effectiveness of the Tail-Drop schedule (achieving optimal FAD and Onset F1 at $\mathcal{I}=[0.00, 0.60]$ for MTM generation, as well as superior FSD and competitive WER at $\mathcal{I}=[0.00, 0.80]$ for English/Chinese TTS). Additionally, omitting guidance during the initial steps causes marked performance degradation across both tasks. Finally, these findings directly align with our proposed scheduling principle for Transcription-to-Audio systems.

\section{Conclusion}

P-MUSE is one MIDI-to-Music model for instrumental music generation and local editing with prompt-MIDI-optional inputs. We design a multi-stage Curriculum-Learning schedule for different prompt settings adaptation and construct the model with fill-in-the-middle formulation casting generation and local editing as the same context-infilling task. Additionally, our Tail-Drop strategy provides a generalized phase-aware CFG schedule for Transcription-to-Audio systems, simultaneously boosting generation quality and inference efficiency while preserving timbre and transcription control.
We further construct the first benchmark spanning paired, style, and mixed prompt settings, with separate generation and editing tasks across piano, guitar, bass, and drums.
Despite its strong performance, P-MUSE does not explicitly incorporate holistic consistency losses for local editing during training. Furthermore, the generated music occasionally lacks expressive control on performance techniques and dynamics in complex performance passages, resulting in slight mechanical artifacts. We leave these to future works.



\bibliography{ref}


\newpage
\subsection{Appendix A: Derivation of Closed-Form Conditional Velocity Fields}

This section derives the closed-form velocity fields used in the main paper. We follow the linear Gaussian path
\begin{equation}
    x_t=\alpha_t x_1+\sigma_t x_0,\qquad x_0\sim\mathcal{N}(0,I),
\end{equation}
where $x_0$ is independent of $x_1$. Solving the path equation for $x_0$ gives
\begin{equation}
    x_0=\frac{x_t-\alpha_t x_1}{\sigma_t}
\end{equation}
Since the CFM conditional velocity is $u_t(x_t\mid x_1)=\dot{\alpha}_t x_1+\dot{\sigma}_t x_0$, substituting the expression above yields
\begin{equation}
\begin{array}{rcl}
u_t(x_t\mid x_1)
&=&\dot{\alpha}_t x_1
  +\dot{\sigma}_t\frac{x_t-\alpha_t x_1}{\sigma_t} \\
&=&\left(\dot{\alpha}_t
  -\frac{\alpha_t\dot{\sigma}_t}{\sigma_t}\right)x_1
  +\frac{\dot{\sigma}_t}{\sigma_t}x_t  \\
&=&A_t x_1+B_t x_t
\end{array}
\end{equation}
where
\begin{equation}
    A_t=\dot{\alpha}_t-\frac{\alpha_t\dot{\sigma}_t}{\sigma_t}
    \qquad
    B_t=\frac{\dot{\sigma}_t}{\sigma_t}
\end{equation}
Taking the conditional expectation given $x_t$ gives
\begin{equation}
\begin{array}{rcl}
u_t^\star(x_t,t)
&:=&\mathrm{E}\!\left[u_t(x_t\mid x_1)\mid x_t\right] \\
&=&A_t\mathrm{E}\!\left[x_1\mid x_t\right]+B_t x_t 
\end{array}
\label{eq:expectation_of_ut}
\end{equation}

$\mathrm{E}[x_1\mid x_t]$ can be estimated by Nadaraya-Watson estimator~\cite{nadaraya1964estimating}. Let the data distribution be approximated by the empirical distribution over $\mathcal{D}=\{x_1^i\}_{i=1}^{N}$:
\begin{equation}
    p_{\mathrm{data}}(x_1)\approx
    \frac{1}{N}\sum_{i=1}^{N}\delta(x_1-x_1^i)
\end{equation}
Conditioning on sample $x_1^i$ makes the path marginal Gaussian,
\begin{equation}
    p_t(x_t\mid x_1^i)=
    \frac{1}{(2\pi\sigma_t^2)^{d/2}}
    \exp\!\left(
    -\frac{\|x_t-\alpha_t x_1^i\|_2^2}{2\sigma_t^2}
    \right)
\end{equation}
For compactness, define
\begin{equation}
    \ell_i(x_t,t)=
    \exp\!\left(
    -\frac{\|x_t-\alpha_t x_1^i\|_2^2}{2\sigma_t^2}
    \right)
\end{equation}
Bayes' rule gives the posterior mass of endpoint $x_1^i$ as
\begin{equation}
\begin{array}{rcl}
P(x_1=x_1^i\mid x_t)
&=&\displaystyle
\frac{p_t(x_t\mid x_1^i)P(x_1=x_1^i)}
{\sum_{j=1}^{N}p_t(x_t\mid x_1^j)P(x_1=x_1^j)}
\\[8pt]
&=&\displaystyle
\frac{(2\pi\sigma_t^2)^{-d/2}\ell_i(x_t,t)\cdot\frac{1}{N}}
{\sum_{j=1}^{N}(2\pi\sigma_t^2)^{-d/2}
\ell_j(x_t,t)\cdot\frac{1}{N}}
\\[8pt]
&=&\displaystyle
\frac{\ell_i(x_t,t)}{\sum_{j=1}^{N}\ell_j(x_t,t)}
=\gamma_i^{\emptyset}(x_t,t)
\end{array}
\end{equation}
Thus, after the shared Gaussian normalization and the uniform prior $1/N$ cancel, the posterior weight is
\begin{equation}
\gamma_i^{\emptyset}(x_t,t)=
\frac{
\exp\!\left(-\frac{\|x_t-\alpha_t x_1^i\|_2^2}{2\sigma_t^2}\right)}
{\sum_{j=1}^{N}
\exp\!\left(-\frac{\|x_t-\alpha_t x_1^j\|_2^2}{2\sigma_t^2}\right)} 
\end{equation}
Therefore,
\begin{equation}
    \mathrm{E}[x_1\mid x_t]
    =\sum_{i=1}^{N}\gamma_i^{\emptyset}(x_t,t)x_1^i
\end{equation}
and substituting this posterior mean into Eq.~\ref{eq:expectation_of_ut} gives
\begin{equation}
    u_{\mathrm{uncond}}^\star(x_t,t)
    =A_t\sum_{i=1}^{N}\gamma_i^{\emptyset}(x_t,t)x_1^i+B_t x_t 
\end{equation}
For the rectified schedule $(\alpha_t,\sigma_t)=(t,1-t)$, we have $A_t=1/(1-t)$ and $B_t=-1/(1-t)$, hence
\begin{equation}
    u_{\mathrm{uncond}}^\star(x_t,t)
    =
    \frac{\sum_{i=1}^{N}\gamma_i^{\emptyset}(x_t,t)x_1^i-x_t}{1-t}
\end{equation}
with
\begin{equation}
\gamma_i^{\emptyset}(x_t,t)=
\frac{
\exp\!\left(-\frac{\|x_t-tx_1^i\|_2^2}{2(1-t)^2}\right)}
{\sum_{j=1}^{N}
\exp\!\left(-\frac{\|x_t-tx_1^j\|_2^2}{2(1-t)^2}\right)} 
\end{equation}
which is the unconditional closed form of the velocity field stated in the main paper.

We now consider the velocity field under external condition $c$. Eq.~\ref{eq:expectation_of_ut} thus yields
\begin{equation}
\begin{array}{rcl}
u_t^\star(x_t,t,c)
&:=&\mathrm{E}\!\left[u_t(x_t\mid x_1)\mid x_t,c\right] \\
&=&A_t\mathrm{E}\!\left[x_1\mid x_t,c\right]+B_t x_t 
\end{array}
\label{eq:appendix_conditional_identity}
\end{equation}
Let $\pi_i(c)=P(I=i\mid C=c)$ denote the conditional empirical prior over the finite dataset $\{(x_1^i,c_i)\}_{i=1}^{N}$. Bayes' rule gives
\begin{equation}
\gamma_i(x_t,t,c)=
\frac{
\pi_i(c)
\exp\!\left(-\frac{\|x_t-\alpha_t x_1^i\|_2^2}{2\sigma_t^2}\right)}
{\sum_{j=1}^{N}
\pi_j(c)
\exp\!\left(-\frac{\|x_t-\alpha_t x_1^j\|_2^2}{2\sigma_t^2}\right)} 
\label{eq:conditional_posterior}
\end{equation}
so that
\begin{equation}
    \mathrm{E}[x_1\mid x_t,c]
    =\sum_{i=1}^{N}\gamma_i(x_t,t,c)x_1^i 
\end{equation}
Combining this posterior mean with Eq.~\ref{eq:appendix_conditional_identity} gives the general conditional velocity field
\begin{equation}
    u_{\mathrm{cond}}^\star(x_t,t,c)
    =A_t\sum_{i=1}^{N}\gamma_i(x_t,t,c)x_1^i+B_t x_t 
\label{eq:general_conditional_velocity_field}
\end{equation}
Under the rectified schedule, it becomes
\begin{equation}
    u_{\mathrm{cond}}^\star(x_t,t,c)
    =
    \frac{\sum_{i=1}^{N}\gamma_i(x_t,t,c)x_1^i-x_t}{1-t}
\label{eq:rectified_conditional_velocity_field}
\end{equation}

For a class-level condition, let $\mathcal{I}_c=\{i:c_i=c\}$ be the selected subset. The conditional empirical prior is $\pi_i(c)=1/|\mathcal{I}_c|$ for $i\in\mathcal{I}_c$ and $\pi_i(c)=0$ otherwise. The common factor $1/|\mathcal{I}_c|$ cancels in Eq.~\ref{eq:conditional_posterior}, yielding
\begin{equation}
\gamma_i^c(x_t,t)=
\frac{
\exp\!\left(-\frac{\|x_t-tx_1^i\|_2^2}{2(1-t)^2}\right)}
{\sum_{j\in\mathcal{I}_c}
\exp\!\left(-\frac{\|x_t-tx_1^j\|_2^2}{2(1-t)^2}\right)}
\end{equation}
for $i\in\mathcal{I}_c$. Substituting these weights into Eq.~\ref{eq:rectified_conditional_velocity_field} gives
\begin{equation}
    u_{\mathrm{cond}}^\star(x_t,t,c)
    =
    \frac{\sum_{i\in\mathcal{I}_c}\gamma_i^c(x_t,t)x_1^i-x_t}{1-t}
\end{equation}

For an instance-level condition, the condition uniquely identifies one sample $x_1^{k(c)}$. In this case, $\pi_i(c)=\mathbf{1}[i=k(c)]$, and Eq.~\ref{eq:conditional_posterior} gives $\gamma_i(x_t,t,c)=\mathbf{1}[i=k(c)]$ for all $t$. Therefore, Eq.~\ref{eq:rectified_conditional_velocity_field} reduces to
\begin{equation}
    u_{\mathrm{cond}}^\star(x_t,t,c)
    =
    \frac{x_1^{k(c)}-x_t}{1-t}
\end{equation}
Since the rectified path satisfies $x_t=(1-t)x_0+t x_1^{k(c)}$ under this condition, the same field can also be written as
\begin{equation}
    u_{\mathrm{cond}}^\star(x_t,t,c)
    =
    x_1^{k(c)}-x_0
\end{equation}

\subsection{Appendix B: Additional Dataset Details}

\begin{table}[t]
    \centering
    {\small
    \setlength{\tabcolsep}{4pt}
    \begin{tabular}{@{}lrrr@{}}
    \toprule
    Instrument group & Hours & Timbres & Segments \\
    \midrule
    Bass & 1,969.17 & 40 & 474,061 \\
    Drum & 1,537.27 & 34 & 867,562 \\
    Guitar & 1,073.67 & 25 & 644,000 \\
    Piano & 5,861.66 & 117 & 1,449,904 \\
    \midrule
    Total & 10,441.77 & 216 & 3,435,527 \\
    \bottomrule
    \end{tabular}
    }
    \caption{Final segmented training set after silence filtering and artifact removal.}
    \label{tab:training_data_statistics}
\end{table}

\begin{table}[t]
    \centering
    {\small
    \setlength{\tabcolsep}{4pt}
    \begin{tabular}{@{}lrrr@{}}
    \toprule
    Dataset & Train Timbres & Total Timbres & Held-out Timbres\\
    \midrule
    NSynth & 148 & 166 & 18 \\
    Slakh & 10 & 13 & 3 \\
    SynthTab & 22 & 24 & 2 \\
    e-gmd & 34 & 43 & 9 \\
    \bottomrule
    \end{tabular}
    }
    \caption{Held-out timbre split used for benchmark construction.}
    \label{tab:heldout_timbres}
\end{table}

This section provides additional details on the training dataset and the benchmark construction. P-MUSE uses five open-source datasets with aligned MIDI and audio:
\begin{itemize}
    \item Slakh~\citep{manilow2019slakh} renders MIDI tracks with virtual instruments, preserves separated stems, and mixes them into complete multitrack songs. It contains 2,100 full songs, approximately 145 hours of mixed audio, 34 broad instrument categories, and 187 timbre patches.
    \item GuitarSet~\citep{xi2018guitarset} contains 3 hours of acoustic-guitar recordings performed by six guitarists. It uses hexaphonic pickups to record the six strings as separate channels, which removes cross-string mixture interference and supports precise string-level pitch and timing annotation.
    \item  MAESTRO~\citep{hawthorne2018enabling} is collected from Yamaha Disklavier piano performances, where the instrument records high-precision MIDI while the live performance is captured as audio. It provides more than 200 hours of closely aligned piano audio and MIDI.
    \item SynthTab~\citep{zang2024synthtab} is a large synthetic guitar-oriented corpus generated by rendering MIDI or tablature with virtual instruments and synthesizers, covering 24 timbres and 7,684 hours of audio.
    \item The Expanded Groove MIDI Dataset (e-gmd)~\citep{callender2020egmd} is a large-scale drum dataset with human drummer performances and velocity annotations, containing 444.5 hours.
\end{itemize}

\begin{table*}[t]
    \centering
    {\small
    \setlength{\tabcolsep}{6pt}
    \begin{tabular}{lccc}
    \toprule
    Hyperparameter & Stage 1 & Stage 2 & Stage 3 \\
    \midrule
    Training Dataset & Full dataset (10,442\,h) & High-quality subset (400\,h) & High-quality subset (400\,h) \\
    FIM Modes & All three modes\textdagger & \texttt{prefix-mid} & All three modes\textdagger \\
    MIDI Drop Probability & 0.3 & 0.5 & 0.3 \\
    MIDI Drop Scope & All & All / Prompt & All \\
    Training Steps & 400,000 & 9,000 & 9,000 \\
    Peak Learning Rate & $1 \times 10^{-4}$ & $2 \times 10^{-5}$ & $2 \times 10^{-5}$ \\
    Warmup Steps & 32,000 & 4,000 & 4,000 \\
    \bottomrule
    \end{tabular}}
    \caption{Detailed stage-wise curriculum training hyperparameters for P-MUSE. All three modes\textdagger~include \texttt{prefix-mid-suffix}, \texttt{prefix-mid}, and \texttt{mid-suffix}.}
    \label{tab:supp_stage_hyperparams}
\end{table*}

\begin{table}[t]
    \centering
    {\small
    \setlength{\tabcolsep}{2.0pt}
    \begin{tabular}{@{}l@{\hspace{3.5pt}}lrrr@{}}
    \toprule
    Curriculum & Prompt & FAD $\downarrow$ & Sim. (\%) $\uparrow$ & F1 (\%) $\uparrow$ \\
    \midrule
    S1$\rightarrow$S2$\rightarrow$S3 & Paired & 0.835 & \textbf{93.1}$\pm.2$ & 71.8$\pm.8$ \\
    S1 & Paired & 0.904 & 92.4$\pm.2$ & 70.9$\pm.8$ \\
    S1$\rightarrow$S2 & Paired & 0.839 & 92.6$\pm.2$ & \textbf{73.3}$\pm.8$ \\
    S1$\rightarrow$S3 & Paired & 0.895 & 92.8$\pm.2$ & 72.4$\pm.8$ \\
    Direct & Paired & \textbf{0.763} & 92.8$\pm.2$ & 71.4$\pm.8$ \\
    \addlinespace[1pt]
    \hdashline
    \addlinespace[1pt]
    S1$\rightarrow$S2$\rightarrow$S3 & Style & \textbf{0.808} & \textbf{92.3}$\pm.2$ & 74.4$\pm.7$ \\
    S1$\rightarrow$S2 & Style & 0.852 & 91.7$\pm.2$ & \textbf{75.5}$\pm.7$ \\
    S1$\rightarrow$S3 & Style & 0.834 & 91.7$\pm.2$ & 74.6$\pm.7$ \\
    Direct & Style & 0.871 & 91.7$\pm.2$ & 73.6$\pm.8$ \\
    \addlinespace[1pt]
    \hdashline
    \addlinespace[1pt]
    S1$\rightarrow$S2$\rightarrow$S3 & Mixed & 0.831 & \textbf{93.0}$\pm.2$ & 71.8$\pm.8$ \\
    S1$\rightarrow$S3 & Mixed & 0.893 & 92.6$\pm.2$ & \textbf{72.0}$\pm.8$ \\
    Direct & Mixed & \textbf{0.827} & 92.7$\pm.2$ & 71.1$\pm.8$ \\
    \bottomrule
    \end{tabular}
    }
    \caption{Editing ablation for the multi-stage Curriculum-Learning. F1 denotes onset F1.}
    \label{tab:stage_ablation_editing_appendix}
\end{table}

\begin{table}[t]
    \centering
    {\small
    \setlength{\tabcolsep}{3.5pt}
    \begin{tabular}{@{}lrrr@{}}
    \toprule
    CFG $\mathcal{I}$ & FAD $\downarrow$ & Sim. (\%) $\uparrow$ & Onset F1 (\%) $\uparrow$ \\
    \midrule
    full & 0.835 & \textbf{93.1}$\pm.2$ & 71.8$\pm.8$ \\
    $[0.00,0.60]$ & \textbf{0.638} & 93.0$\pm.2$ & \textbf{72.9}$\pm.8$ \\
    $[0.00,0.80]$ & 0.668 & 93.0$\pm.2$ & 72.7$\pm.8$ \\
    $[0.00,0.92]$ & 0.699 & \textbf{93.1}$\pm.2$ & 72.5$\pm.8$ \\
    $[0.04,1.00]$ & 0.769 & 92.3$\pm.2$ & 71.0$\pm.8$ \\
    $[0.08,1.00]$ & 1.060 & 91.2$\pm.2$ & 70.1$\pm.8$ \\
    \bottomrule
    \end{tabular}
    }
    \caption{Additional paired-prompt editing ablations of Tail-Drop on P-MUSE.}
    \label{tab:tail_drop_grid_paired_editing}
\end{table}

\begin{table*}[t]
    \centering
    {\small
    \setlength{\tabcolsep}{3.5pt}
    \begin{tabular}{lcrrr}
    \toprule
    \multicolumn{5}{c}{(a) Style prompt} \\
    \midrule
    Task & CFG $\mathcal{I}$ & FAD $\downarrow$ & Sim. (\%)$\uparrow$ & Onset F1 (\%)$\uparrow$ \\
    \midrule
    Gen. & full & 0.558 & \textbf{92.9}$\pm.5$ & 73.0$\pm1.9$ \\
    Gen. & $[0.00,0.60]$ & \textbf{0.468} & \textbf{92.9}$\pm.4$ & 72.9$\pm1.9$ \\
    Gen. & $[0.00,0.80]$ & 0.480 & \textbf{92.9}$\pm.5$ & \textbf{73.8}$\pm1.8$ \\
    Gen. & $[0.00,0.92]$ & 0.491 & \textbf{92.9}$\pm.5$ & 73.3$\pm1.9$ \\
    Gen. & $[0.04,1.00]$ & 0.655 & 91.1$\pm.6$ & 69.4$\pm2.1$ \\
    Gen. & $[0.08,1.00]$ & 1.780 & 89.2$\pm.6$ & 63.0$\pm2.2$ \\
    \addlinespace[1pt]
    \hdashline
    \addlinespace[1pt]
    Edit & full & 0.808 & \textbf{92.3}$\pm.2$ & 74.4$\pm.7$ \\
    Edit & $[0.00,0.60]$ & \textbf{0.640} & \textbf{92.3}$\pm.2$ & 74.6$\pm.7$ \\
    Edit & $[0.00,0.80]$ & 0.686 & \textbf{92.3}$\pm.2$ & \textbf{74.9}$\pm.7$ \\
    Edit & $[0.00,0.92]$ & 0.704 & \textbf{92.3}$\pm.2$ & 74.8$\pm.7$ \\
    Edit & $[0.04,1.00]$ & 0.750 & 91.4$\pm.2$ & 70.9$\pm.8$ \\
    Edit & $[0.08,1.00]$ & 1.280 & 90.0$\pm.2$ & 64.1$\pm.9$ \\
    \bottomrule
    \end{tabular}
    \hspace{12pt} 
    \begin{tabular}{lcrrr}
    \toprule
    \multicolumn{5}{c}{(b) Mixed prompt} \\
    \midrule
    Task & CFG $\mathcal{I}$ & FAD $\downarrow$ & Sim. (\%) $\uparrow$ & Onset F1 (\%)$\uparrow$ \\
    \midrule
    Gen. & full & 0.668 & \textbf{94.2}$\pm.3$ & 74.2$\pm1.9$ \\
    Gen. & $[0.00,0.60]$ & \textbf{0.542} & 94.1$\pm.4$ & 74.6$\pm1.9$ \\
    Gen. & $[0.00,0.80]$ & 0.579 & 94.0$\pm.4$ & \textbf{74.9}$\pm1.9$ \\
    Gen. & $[0.00,0.92]$ & 0.572 & 94.1$\pm.4$ & 74.7$\pm1.9$ \\
    Gen. & $[0.04,1.00]$ & 0.723 & 92.7$\pm.5$ & 72.7$\pm2.0$ \\
    Gen. & $[0.08,1.00]$ & 1.287 & 90.9$\pm.6$ & 69.5$\pm2.0$ \\
    \addlinespace[1pt]
    \hdashline
    \addlinespace[1pt]
    Edit & full & 0.831 & \textbf{93.0}$\pm.2$ & 71.8$\pm.8$ \\
    Edit & $[0.00,0.60]$ & \textbf{0.615} & 92.9$\pm.2$ & \textbf{72.7}$\pm.8$ \\
    Edit & $[0.00,0.80]$ & 0.670 & \textbf{93.0}$\pm.2$ & 72.5$\pm.8$ \\
    Edit & $[0.00,0.92]$ & 0.697 & \textbf{93.0}$\pm.2$ & 72.2$\pm.8$ \\
    Edit & $[0.04,1.00]$ & 0.788 & 92.3$\pm.2$ & 70.4$\pm.8$ \\
    Edit & $[0.08,1.00]$ & 1.006 & 91.4$\pm.2$ & 69.5$\pm.8$ \\
    \bottomrule
    \end{tabular}
    }
    \caption{Additional style- and mixed-prompt ablations of Tail-Drop on P-MUSE. (a) Style prompt; (b) Mixed prompt.}
    \label{tab:tail_drop_grid_style_mixed}
\end{table*}

We additionally render MIDI performances with NSynth~\citep{engel2017nsynth} as a sample-based synthesizer.
The original NSynth dataset contains acoustic, electronic, and synthetic timbre categories. We use only the acoustic and electronic categories, and separate single-track MIDI files from the Lakh MIDI Dataset~\citep{raffel2016learning} according to General MIDI rules.
Because NSynth provides only discrete velocity levels, each MIDI velocity is quantized to its nearest available value in $\{25,50,75,100,127\}$. The matched NSynth recording with the same pitch and quantized velocity is then placed at the MIDI note onset, modulated by the preset envelope and mixed into the output waveform. To support notes longer than the original sample, the sustain portion is extended by looping a fixed segment with cross-fading, followed by attack and release envelopes to reduce artifacts. Rendered clips with missing pitch or velocity samples are discarded.
In the NSynth rendering pipeline, we render bass and piano tracks (with keyboards included under piano), alongside a few guitar tracks. Crucially, the rendered guitar audio is excluded from training and reserved exclusively for out-of-distribution (OOD) benchmark testing, as note-based rendering lacks expressive performance dynamics and is thus ill-suited for training. For training, this rendering pipeline produces 36 bass timbres, including 1 acoustic and 35 electronic timbres, and 112 piano timbres, including 18 acoustic and 94 electronic timbres.

Full recordings are segmented into 3--30\,s clips, with their metadata recorded in a JSONL file. We apply voice-activity detection to avoid silent segments. Since Slakh and SynthTab are synthetic corpora and have been reported by the research and open-source communities to contain occasional sound-leakage artifacts, we remove the affected segments during pre-processing. The final training dataset as shown in Table~\ref{tab:training_data_statistics} contains 10,441.77 hours, 216 timbres, and 3,435,527 audio/MIDI segments in total, averaging approximately 50 hours per timbre after segmentation.

\subsection{Appendix c: Additional Benchmark Details}
The benchmark is constructed from timbres held out from training (Table~\ref{tab:heldout_timbres}), as well as some guitar segments rendered using NSynth. We construct three benchmarks for the paired, style, and mixed prompt settings. Each benchmark independently samples 3--30\,s target audio/MIDI clips from the held-out timbres, with 100 clips per instrument group. Specifically, the paired and style prompt benchmarks share an identical composition, differing only in whether prompt MIDI is loaded at inference time. The mixed prompt benchmark adopts the same segments as the former two benchmarks but with different slicing intervals, supplying both prompt types simultaneously.
Each target clip defines one generation case and five local-editing cases: note addition, note deletion, pitch shift, velocity scaling, and timing perturbation. Specifically, generation follows a \texttt{prefix-mid} mode, whereas local editing adopts the standard \texttt{prefix-mid-suffix} mode. Therefore, each instrument group contributes 600 cases per benchmark variant, and each variant contains 2,400 cases across bass, drum, guitar, and piano.

\subsection{Appendix D: Implementation Details}

\paragraph{Architecture Details.}
P-MUSE contains 481.3M parameters in total. The generative backbone is a Diffusion Transformer (DiT)~\citep{peebles2023dit} with a hidden dimension of 1024, 25 layers, 16 attention heads (head dimension 64). We condition the backbone on 128-dimensional mel latents and 512-dimensional MIDI tokens. We optimize the standard DiT block with rotary positional embeddings~\citep{su2024roformer}, RMS normalization~\citep{zhang2019rmsnorm}, and QK-RMSNorm~\citep{henry2020qknorm}. During training, time steps are sampled from a logit-normal distribution. Following the $x$-prediction, $v$-loss configuration of JiT~\citep{li2025backtobasics}, the DiT estimates the clean mel endpoint $\hat{x}_{1,\theta}$ and derives $v_\theta(x_t,t,h)=(\hat{x}_{1,\theta}-x_t)/(1-t)$. The model is optimized using an $L_2$ loss between $v_{\mathrm{pred}}$ and $v_{\mathrm{gt}}$, applied exclusively to the target region.

\paragraph{Training Details.}
We optimize all stages using AdamW~\cite{loshchilov2019decoupledweightdecayregularization} ($\beta_1 = 0.9$, $\beta_2 = 0.999$, $\epsilon = 10^{-8}$, weight decay $0.01$) with gradient clipping set to $0.2$ and a fixed random seed of 114. We employ an inverse-square-root learning-rate schedule across all stages. For Stage~1 (S1 ICL), training runs for up to 400,000 steps with a 32,000-step linear warmup at a peak learning rate of $1 \times 10^{-4}$. For Stage~2 (S2 Prefix CFG) and Stage~3 (S3 Unified ICL), the peak learning rate is adjusted to $2 \times 10^{-5}$ with a 4,000-step linear warmup for up to 9,000 steps. Training is conducted on 8 GPUs with a batch size of 7 per GPU.

\paragraph{Curriculum-Learning Pipeline.}
For the SFT stages (Stage 2 and Stage 3), we rank training clips using the AudioBox evaluation model~\citep{vyas2023audioboxunifiedaudiogeneration} and retain a 400-hour high-quality subset, comprising 100 hours each of piano, guitar, bass, and drums.
Stage~1 is trained on the full training dataset across three fill-in-the-middle (FIM) modes (\texttt{prefix-mid-suffix}, \texttt{prefix-mid}, and \texttt{mid-suffix}) with a global MIDI drop probability of 0.3. Stage~2 focuses on adapting to style prompting using the high-quality subset under the \texttt{prefix-mid} mode, with a 0.5 MIDI drop probability which is then equally split between dropping all MIDI and prompt MIDI. Stage~3 unifies generation and editing by prepending style prompts while restoring all three FIM infilling modes. The detailed stage-wise curriculum configuration is summarized in Table~\ref{tab:supp_stage_hyperparams}.

\paragraph{Inference Details.}
At inference, we integrate the reverse ODE using a 25-step first-order Euler solver and decode the predicted mel-spectrograms with a pre-trained Vocos vocoder~\citep{siuzdak2024vocosclosinggaptimedomain}. For local editing, we additionally regenerate all audio frames within a \texttt{release\_duration} window following the edited region to prevent residual sustain of the original audio from interfering with the suffix audio. During inference, \texttt{release\_duration} is set to 1\,s for the paired and mixed prompt settings, whereas it is set to 0\,s for the style prompt setting. 
Unless otherwise stated, all benchmark results use the checkpoint after multi-stage Curriculum-Learning and CFG ($s_{\mathrm{cfg}}=2.0$) applied at every integration step.

\subsection{Appendix E: Additional Editing Ablations}

Table~\ref{tab:tail_drop_grid_paired_editing} reports the result of the ablations of Tail-Drop strategy on paired-prompt editing and Table~\ref{tab:tail_drop_grid_style_mixed} reports the generation and editing under style and mixed prompt settings. 
Across prompt settings and tasks, applying CFG in the early interval and switching to the unconditional field near the data endpoint consistently improves FAD over full-trajectory CFG. 
Specifically, the interval $\mathcal{I}=[0.00,0.60]$ achieves the best FAD across all reported groups while maintaining stable and highly competitive timbre similarity and onset F1 scores. In contrast, omitting early guidance (e.g., $\mathcal{I}=[0.04,1.00]$ and $\mathcal{I}=[0.08,1.00]$) substantially degrades performance across all three metrics. These empirical results further support the theoretical insights and practical effectiveness of the proposed phase-aware CFG scheduling principle for Transcription-to-Audio systems like MTM and TTS and Tail-Drop strategy.

\end{document}